# Dynamic Nuclear Polarization Mechanisms using TEMPOL and trityl OX063 radicals at 1 T and 77 K

Ewoud Vaneeckhaute*[1], Charlotte Bocquelet[1], Nathan Rougier[1], Shebha Anandhi Jegadeesan[2], Sanjay Vinod-Kumar[2], Guinevere Mathies[2], Roberto Melzi[3], James Kempf[4], Quentin Stern[1], Sami Jannin[1]

A sensitivity increase of two orders of magnitude in proton ($^1$H) and carbon ($^{13}$C) spins via dynamic nuclear polarization (DNP) has been accomplished recently using a compact benchtop DNP polarizer operating at 1 T and 77 K. However the DNP mechanisms at play at such low magnetic field and high operating temperature are still not elucidated. A deeper understanding of the dominant polarization transfer mechanisms between electrons and nuclear spins at these unconventional benchtop conditions is therefore required if one wants to devise strategies to boost sensitivity further. In this study, we found that DNP is generally dominated by solid effect for narrow electron paramagnetic resonance (EPR) line radicals (15 mM trityl OX063) and cross effect for broad EPR line radicals (50 mM TEMPOL). For both radicals, the dominant DNP mechanisms were investigated varying the microwave frequency and measuring the $^1$H and $^{13}$C DNP enhancement factors to obtain $^1$H and $^{13}$C DNP spectra. The impact of varying the microwave power on the $^1$H DNP buildup times and the $^1$H nuclear spin relaxation times were important as well to distinguish between solid effect and cross effect DNP. Finally, time-resolved electron saturation simulations under continuous microwave irradiation could replicate the experimental $^1$H and $^{13}$C DNP spectra at 1 T and 77 K for both radicals considering their electron relaxation properties. Only for trityl OX063, the $^{13}$C DNP spectra showed additional DNP maxima compared to the simulations. This has been attributed to methyl rotor induced $^1$H-$^{13}$C heteronuclear cross relaxation in [1-$^{13}$C] acetate present at 1 T and 77 K.

## 1. Introduction

Boosting nuclear magnetic resonance (NMR) sensitivity in harmony with its general applicability, high resolution and non-destructiveness remains a major scientific challenge.[1] Dynamic nuclear polarization (DNP) emerged as a front runner in the development towards tackling this challenge. With DNP, analytes brought in close contact with a polarizing agent (e.g. TEMPOL or trityl) — a molecule with unpaired electrons — in a frozen and glassy state can become repetitively fuelled with hyperpolarization through microwave (μw) irradiation.[2,3] Although DNP, compared to many other hyperpolarization strategies[4–6], does not compromise on general applicability, it still suffers from low resolution when measurements are done in the frozen state, due to the presence of strong dipolar interactions (kHz) that obscure the fine structural features (Hz) at the heart of structural analysis. An elegant, but destructive solution to allow high-resolution analysis, is dissolution DNP (dDNP) which induces a solid-to-liquid phase transition by rapid dissolution and dilution of the sample with a hot and pressurized solvent.[7] This provides access to liquid-state NMR detection where strong dipolar interactions (kHz) are averaged out.[8] Despite these advancements, the dissolution process used for the solid-to-liquid transitions remains intrinsically destructive and thus irreversible. The hyperpolarized signal can therefore be manipulated and detected only once, briefly after the dissolution. A key requirement to expand the application potential of dDNP could therefore be to find a recyclable DNP alternative.[9–11] We aim to reach a new horizon of DNP applications by enabling a repeated sequence of solid-state DNP hyperpolarization followed by dilution-free melting, so that multiple solution-state NMR observations are made possible on the same analyte. This aims at creating compatibility with NMR experiments that require numerous consecutive acquisitions for either phase cycling or multidimensional detection schemes without comprising on general applicability, high resolution, and non-destructiveness, the key strengths of NMR.[12]

To pursue this ambitious goal, we based ourselves on the concept of recyclable hyperpolarized flow (HypFlow) DNP[11,13] which combines the use of a compact 1 T benchtop DNP polarizer operating at a temperature of 77 K, together with the use of porous silica-based hyperpolarizing solids (HYPSO) in which paramagnetic nitroxide radicals are immobilized.[14] This opens the way to physically extruding the hyperpolarized analytes out of a heterogenous stationary phase of HYPSO, therefore not contaminated with any glassing agent or solvated paramagnetic species.[15] So far, our benchtop DNP polarizer (Figure 1) succeeded in enhancing the solid-state $^1$H and $^{13}$C NMR signal of a partially protonated frozen solution with 3 M [1-$^{13}$C] sodium acetate up to ε = 100 using free TEMPOL (50 mM) and ε = 62 when using mesoporous hyperpolarizing solids with immobilized TEMPOL (43 μmol.cm$^{-3}$).[13] At these optimal radical concentrations, DNP buildup times for protons were less than 1 second, while relaxation of $^{13}$C spins inside the pores of hyperpolarizing solids could be slowed down considerably up to a $T_1$ of 18 s.[11]

[1.] Université Claude Bernard Lyon 1, CNRS, ENS Lyon, UCBL, CRMN UMR 5082, 69100 Villeurbanne, France
[2.] Department of Chemistry, University of Konstanz, Universitätsstr. 10, 78464, Konstanz, Germany
[3.] Bruker Italia S.r.l., Viale V. Lancetti 43, 20158 Milano, Italy
[4.] Bruker Biospin, Billerica, Massachusetts 01821, United States
† Footnotes relating to the title and/or authors should appear here.

The conditions of 1 T and 77 K in which the benchtop DNP polarizer operates, are however considered unusual from a DNP perspective. In recent years, the pursuit for larger enhancement factors using dDNP has led to an increased focus to polarize at higher magnetic fields (up to 7 T) and lower temperatures (below 2 K).[16,17] At conventional dDNP conditions, TEMPOL has been recognized to effectively hyperpolarize high-gamma nuclei such as protons mainly by means of thermal mixing induced polarization transfer.[18–21] In this case, the presence of a large network of strong dipolar coupled electrons in combination with long spin-lattice electron relaxation times of hundreds of milliseconds makes it possible to efficiently transfer polarization to nearby nuclear spins without requiring strong microwave irradiation. Under such conditions, the spontaneous process of electron spectral spin-diffusion (the spread of polarization between neighboring electron spins and along the EPR line) is fast compared to the spin-lattice electron relaxation ($T_{1S}$)[19] and gives rise to efficient saturation of electrons under microwave irradiation. However, when working at 1 T and 77 K, the difference in magnetic field strength and cryogenic operating temperature impacts the electron and nuclear spin dynamics therefore affecting the DNP dynamics and efficiency considerably. Above 20 K, $T_{1S}$ of TEMPOL quickly decreases to hundreds of microseconds as already shown independently by Shimon et al.[21] and Zhao et al.[22] Consequentially, the question whether other DNP transfer pathways being solid effect (SE) or cross effect (CE) mechanisms are efficient at 1 T and 77 K remains open.

Here, the influence of the microwave frequency and the microwave power on solid-state ¹H and ¹³C DNP performances of broad-line TEMPOL nitroxide radicals (Figure 1a, right) embedded in an amorphous frozen solution with 3 M fully labelled [1-¹³C] sodium acetate are explored and specifically compared to narrow-line trityl OX063 radicals (Figure 1b, left). This way, experimental ¹H and ¹³C DNP spectra, ¹H DNP buildup times, and ¹H spin-lattice relaxation times were measured for both radicals using the benchtop polarizer shown in Figure 1b and Figure 1c. Numerical DNP simulations were then used to describe the interplay and importance of microwave saturation, electron relaxation, and spectral diffusion on the electron saturation efficiency in our specific conditions. Hyperpolarization of the nuclear spins was then estimated by discretizing the EPR lineshape into electron bins which can exchange polarization to nearby nuclear spins depending on the required matching conditions for SE or CE in analogy with the methodology of Hovav et al.[23]

By comparison of the DNP simulations with experimental ¹H and ¹³C DNP spectra, we were able to unravel the DNP mechanisms at play. For TEMPOL at 50 mM, CE is dominant for ¹H and ¹³C, and the shape of the ¹H and ¹³C DNP spectra depicted could be well reproduced. Yet, surprisingly also for 15 mM trityl OX063, besides SE DNP, polarization transfer to protons caused by CE is present at 1 T and fits well with the appearance of a broad component in the ¹H DNP spectrum. For ¹³C in case of trityl, heteronuclear cross relaxation between protons and carbons of the methyl rotor in acetate are causing additional DNP extrema in the ¹³C DNP spectrum at the position where ¹H DNP is best.

## 2. Results and discussion

### 2.1 Effect of microwave frequency on ¹H and ¹³C DNP enhancement

A DNP spectrum plots the enhancement of hyperpolarized nuclear spins against the microwave irradiation frequency ($\omega_m$) and provides a first access point to which DNP mechanisms could be dominating. For this reason, experimental ¹H and ¹³C DNP spectra for 15 mM trityl OX063 (Figure 1a, left) and 50 mM TEMPOL (Figure 1a, right) were obtained with the benchtop polarizer shown in Figure 1b and Figure 1c. The pulse trains used for obtaining the ¹H and ¹³C DNP spectra are shown in respectively Figure 6a and Figure 6b in the materials and methods section. The resulting DNP spectra are shown in Figure 2. The sample for acquiring the ¹H DNP spectra consisted of the radicals solvated in 2:2:6 H₂O:D₂O:DMSO-d₆ ($v$:$v$:$v$) solutions. The solutions were then frozen and measured in the 1 T benchtop DNP polarizer operating at 77 K. For the ¹³C DNP spectra, the same sample composition as before was used with 3 M fully labelled [1-¹³C] sodium acetate added. More detailed information on the sample preparation, benchtop polarizer, NMR acquisitions, and processing workflow can be found in the materials and methods section.

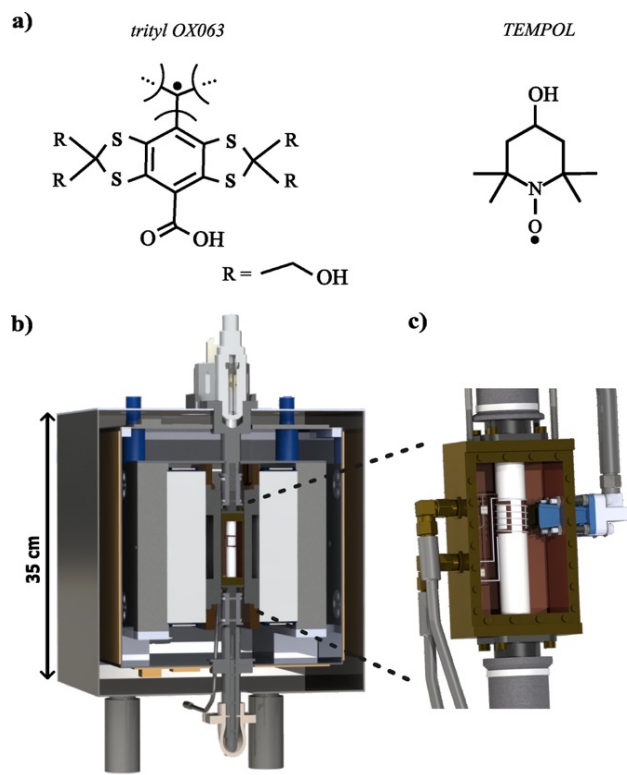

**Figure 1. The polarizing agents and benchtop polarizer used for studying solid-state dynamic nuclear polarization (DNP) mechanisms at 1 T and 77 K**. a) The Lewis structure of trityl OX063, a typical narrow EPR line DNP polarization agent with an unpaired electron on the central carbon, and of TEMPOL, a typical broad EPR line DNP polarization agent with an unpaired electron on the nitroxide functionality. b) Frontal cut of the prototype benchtop 1 T DNP polarizer co-developed with Bruker Biospin. c) DNP-NMR cavity equipped with a ¹³C solenoid radio-frequency (rf) coil and a ¹H saddle rf coil tuned externally to 10.7 MHz and 42.57 MHz respectively. A rectangular antenna horn directs the microwaves onto the sample location after the coaxial-to-waveguide transition from the Kα-band microwave source (26.6 to 28.8 GHz up to 5 watts).

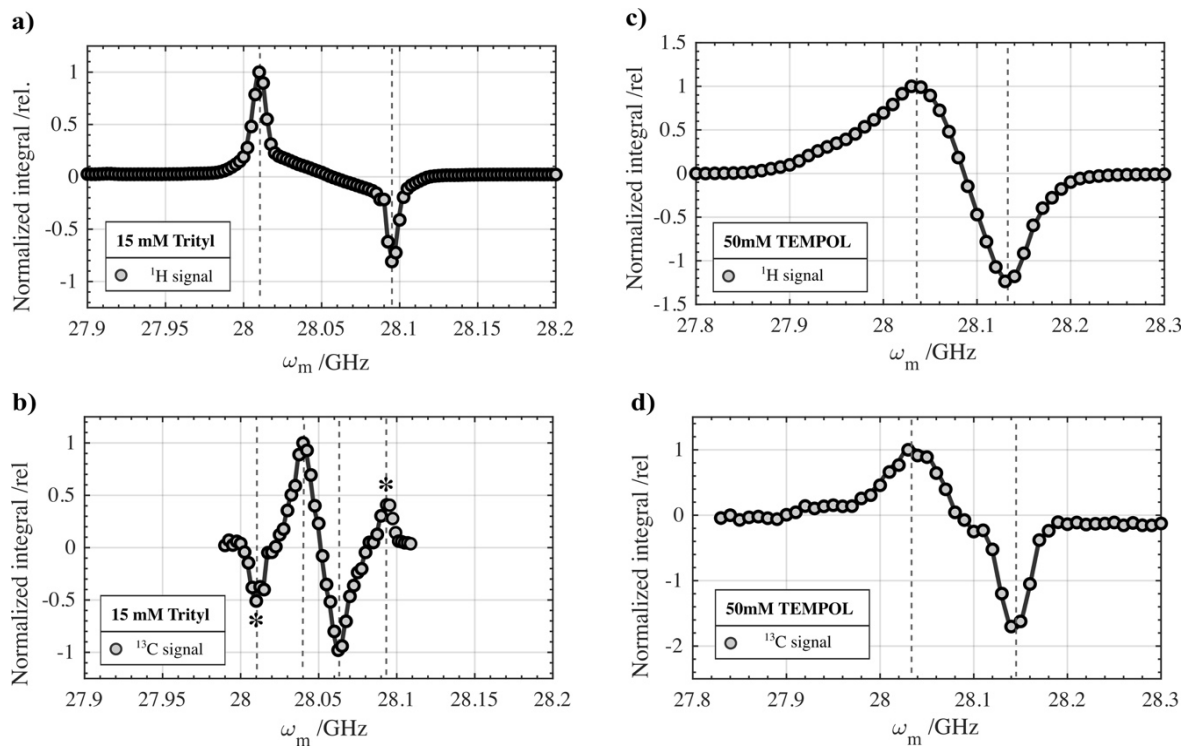

**Figure 2. The effect of microwave frequency on the $^1$H and $^{13}$C DNP enhancement measured at 1 T and 77 K using trityl OX063 (left) and TEMPOL (right).** $^1$H and $^{13}$C DNP spectra show the relative proton and carbon nuclear spin hyperpolarization versus microwave frequency ($\omega_S$) obtained using the benchtop DNP polarizer described in Figure 1. **a)** $^1$H DNP spectrum of a 2:2:6 H$_2$O:D$_2$O:DMSO-d$_6$ (v:v:v) solution using 15 mM trityl OX063. **b)** $^{13}$C DNP spectrum of the same DNP sample composition but with 3M fully labelled [1-$^{13}$C] sodium acetate added. The asterisks indicate the appearance of additional DNP optima with opposite polarization assigned to $^1$H-$^{13}$C heteronuclear cross relaxation in acetate. **c)** $^1$H DNP spectrum of a 2:2:6 H$_2$O:D$_2$O:DMSO-d$_6$ (v:v:v) solution using 50 mM TEMPOL. **d)** $^{13}$C DNP spectrum of the same DNP sample composition using TEMPOL with 3 M fully labelled [1-$^{13}$C] sodium acetate added.

In case of trityl OX063, the $^1$H and $^{13}$C DNP spectra shown in Figure 2a,b are in agreement with SE dominated DNP at 1 T and 77 K. In case of solid effect, microwave irradiation with frequency $\omega_m$ drives polarization transfer from a single electron to nearby nuclear spins through allowing second-order electron-nuclear transitions.[24–26] These transitions are shifted by the Larmor frequency of the nuclear spin ($\omega_I$) at $\omega_S + \omega_I$ and $\omega_S - \omega_I$ in a 1 electron (e) – 1 nucleus (n) two-spin system. The frequency difference for (+) and (-) DNP optima observed at respectively 28.01 GHz and 28.095 GHz corresponds to 85 MHz and is therefore two times $\omega_I^{1H}$ of 42.57 MHz at 1 T. When hyperpolarization is monitored for $^{13}$C in Figure 2b, the main DNP optima shift closer together separated by 21.5 MHz again corresponding to two times $\omega_I^{13C}$ of 10.73 MHz. Unexpectedly, additional optima with opposite sign (indicated with an asterisk in Figure 2b) appeared and are attributed to heteronuclear cross relaxation between protons and carbons of the methyl rotor in acetate, as discussed later (*vide infra*).[27,28] In contrast, for TEMPOL the $^1$H and $^{13}$C DNP spectra shown respectively in Figure 2c and Figure 2d are characterized by a much broader appearance. Here, the frequency difference between the (+) and (-) DNP optima is ~100 MHz independent on the nuclear spin under observation. This complicates the analysis and does not give any clear indication on the contribution of SE or CE DNP. Other experimental methods were therefore explored to provide insights into the dominant polarization transfer mechanisms occurring at 1 T and 77 K.

## 2.2 Effect of microwave power on $^1$H DNP enhancements and $^1$H DNP buildup times

The DNP enhancement factor $\varepsilon_{DNP}$ and the DNP buildup time $\tau_{DNP}$ of bulk protons in frozen DNP solution were monitored by changing the microwave power ($P_{\mu w}$) on both polarizing agents. Practically, $\tau_{DNP}$ was monitored by using consecutive $\pi/36$ small-angle radio-frequency (rf) pulses with a predefined delay while the microwave source was on. The pulse train can be found in Figure 6c in the materials and methods section. Then, after gating the microwave, the polarization decay was monitored analogously to estimate the bulk proton relaxation time ($T_1$). On the left and right, in Figure 3a and Figure 3d, the hyperpolarization buildup and decay curves of respectively trityl OX063 and TEMPOL are plotted for different $P_{\mu w}$. Underneath, the relative enhancement factors $\varepsilon_{DNP}$ are plotted and show the dependency of the maximum signal obtained as function of $P_{\mu w}$ for trityl OX063 in Figure 3b and for TEMPOL in Figure 3e. Also the DNP buildup times $\tau_{DNP}$ and resulting nuclear spin relaxation times $T_1$ values were extracted by fitting with the exponential buildup and decay functions of respectively equations 13 and 14. The results are shown for trityl OX063 and TEMPOL in respectively Figure 3c and Figure 3f.

For trityl OX063, the relative enhancement factor $\varepsilon_{DNP}$ is linearly proportional to $P_{\mu w}$. The maximum DNP enhancement that can be reached is limited by the available microwave power of the polarizer. The buildup time $\tau_{DNP}$ gradually increases and starts differing from $T_1$ at a sufficiently high value of $P_{\mu w}$.[26] Overall $\tau_{DNP}$ changes from around 14 s ($\tau_{DNP} = T_1$) for 0.55 W to 11 s ($\tau_{DNP} < T_1$) for 4.57 W microwave output power. In contrast, for TEMPOL, $\varepsilon_{DNP}$ approaches an asymptote as shown in Figure 3e and $\tau_{DNP}$ is independent of the microwave power, remaining unchanged at 0.4 s as shown in Figure 3f.

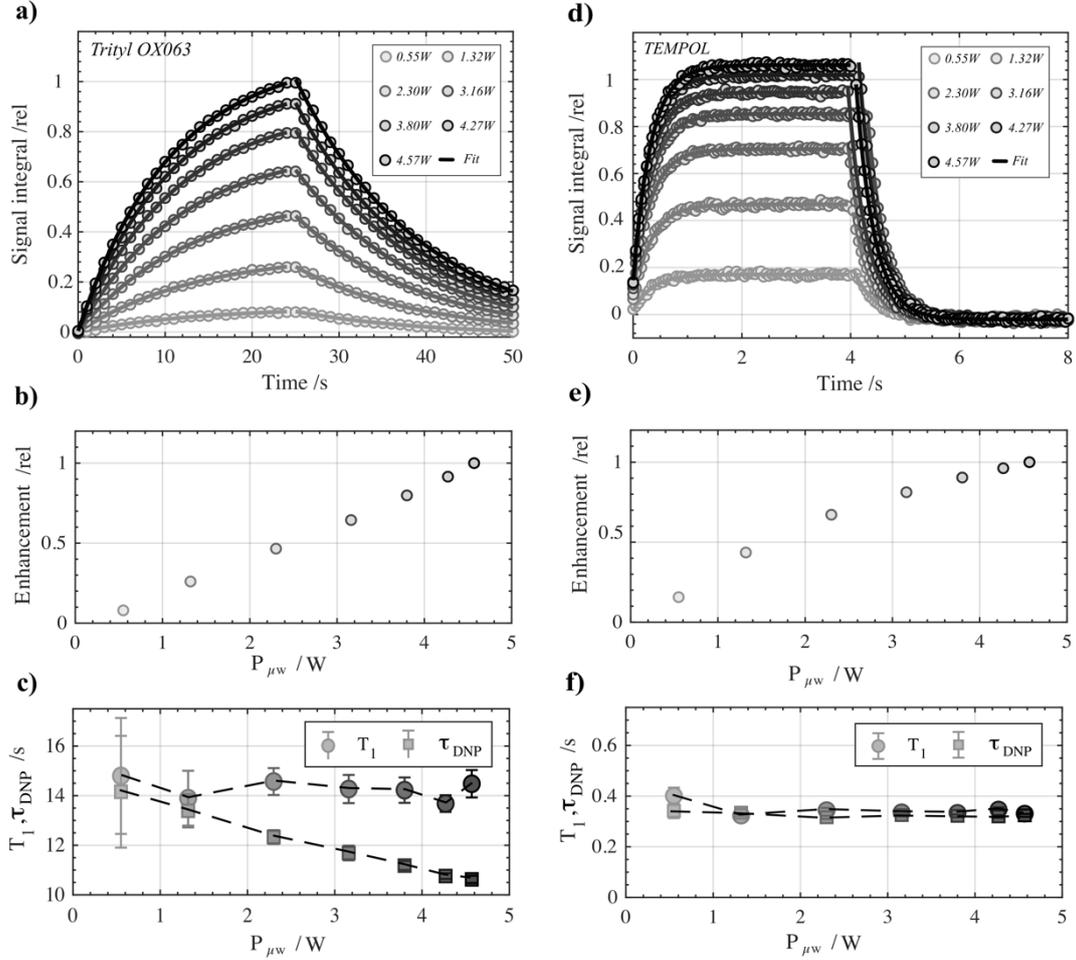

**Figure 3. The effect of microwave power on the ¹H DNP buildup time (μw on) and ¹H bulk spin lattice relaxation time (μw off) and the normalized ¹H DNP enhancement at 1 T and 77 K using trityl OX063 (left) and TEMPOL (right) at 1 T and 77 K.** The characteristic buildup and relaxation times were measured under varying microwave power and frequency in a 2:2:6 H₂O:D₂O:DMSO-d$_6$ (v:v:v) solution. The microwave source (28 GHz) capable of varying the microwave power from 0 → 5 W and the benchtop DNP polarizer (42.57 MHz) described in Figure 1 were used for analysis. **a,d)** ¹H DNP buildup and decay monitored via 5 degree low-angle rf pulses at various microwave power of 0.55, 1.32, 2.3, 3.16, 3.8, 4.27, 4.57 W at 28.095 GHz for trityl OX063 and 28.13 GHz for TEMPOL. **b,e)** Normalized DNP enhancement extracted from the buildup curves plotted for the different microwave power. **c,f)** Fitted ¹H characteristic buildup and relaxation times at different microwave powers for trityl OX063 and TEMPOL. All data were treated using MATLAB and fitted with monoexponential functions shown in equations 13 and 14.

The dependency of $\tau_{DNP}$ on $P_{\mu w}$ can be rationalized when looking deeper into which parameters influence polarization transfer in SE and CE DNP. In case that the SE mechanism would be dominant, the microwaves themselves drive polarization transfer as mentioned before. The microwave allows second order zero-quantum (ZQ) and/or double-quantum (DQ) transitions of the e-n two-spin system to exchange populations between partially mixed eigenstates by making them degenerate. In particular, it is the pseudo-secular part of the e-n hyperfine interaction that, during microwave irradiation, is sufficiently strong to allow partial mixing of the Zeeman eigenstates.[25,26] An expression for the polarization exchange state mixing rate $W_{SI}^{\pm}$ that induces either positive (+) or negative (-) hyperpolarization has been derived before[24–26,29] and is proportional to:

$$W_{SI}^{\pm} \propto \frac{|A_{IS}|^2 \omega_{1S}^2}{\omega_I^2}, \quad (1)$$

where $A_{IS}$ is the pseudo-secular e-n hyperfine interaction, $\omega_{1S}$ is the magnetic field strength of the irradiating microwave, and $\omega_I$ is the Larmor frequency of the hyperpolarized nuclear spin. The polarization exchange rate of equation 1 is thus quadratically proportional to the microwave field strength $\omega_{1S}$ which means it is linearly proportional to $P_{\mu w} \propto \omega_{1S}^2$. For trityl OX063, both in Figure 3b and 3c, $P_{\mu w}$ shows a similar influence on the relative enhancement factor $\varepsilon_{DNP}$ and the DNP buildup time $\tau_{DNP}$ of bulk protons. The inverse quadratic dependency of the rate on the Larmor frequency in equation 1 explains the fast ¹H buildup times of 10 s compared to 46 s when similar trityl OX063 concentrations (20 mM) and sample compositions (1:3:6 H₂O:D₂O:DMSO-d$_6$) are used but at higher fields and temperatures (8.9 T, 100K). [30]

In comparison, for TEMPOL, the asymptotic relation $\varepsilon_{DNP}$ on $P_{\mu w}$ and the fact that $\tau_{DNP}$ of bulk protons is independent of the microwave power makes it unlikely that SE could be considered important as a hyperpolarization mechanism when using nitroxide radicals at 50 mM at 1 T and 77 K. On the contrary, CE seems to be a valid explanation for why $\tau_{\text{DNP}}$ equals $T_1$ of the bulk nuclear spins and the rates are independent of the microwave conditions. Instead of second-order ZQ and DQ transitions driven by the microwaves, triple spin flips mediate transfer of polarization imbalance between degenerate eigenstates of neighbouring electrons close to a nuclear spin of interest.[19,31] The triple spin flip polarization exchange rate $W_{\text{SSI}}^{\pm}$ does not depend anymore on the microwave field strength and is proportional to equation 2 disregarding the contribution of exchange coupling interactions:[19,32]

$$W_{\text{SSI}}^{\pm} \propto \frac{|A_{\text{IS}}|^2 D_0^2}{\omega_{\text{I}}^2}, \quad (2)$$

where $D_0$ is the e-e dipolar interaction between two neighboring electrons. Instead, the rate is influenced by the square of the amplitude of $D_0$. In our previous study[13] on nitroxide radical concentrations optimization for benchtop DNP, the bulk $^1$H buildup rate also increased with radical concentration. The relationship between bulk proton buildup rates and radical concentration was however not quadratic but increased with an exponent of 1.58. The reason for this discrepancy is under investigation.

Proceeding with the observation that $T_1$ equals $\tau_{\text{DNP}}$ within the limits of the fit error independent of the microwave power, hints towards the fact that the inherent mechanism for fuelling nuclear spin polarization in the presence of microwaves also must act as an effective relaxation pathway for the nuclear spins in absence of the microwaves. Again, the triple spin flips, central in CE mediated DNP, come forward as a valid explanation. The energy-conservative triple spin flips between two electron spins and a nuclear spin in CE are also an effective cross relaxation pathway for the nuclear spins to exchange energy with the lattice.[33] In that case, bulk $^1$H hyperpolarization buildup times and relaxation times becomes the same as observed empirically (both 0.4 s for 50 mM TEMPOL). Again, the fast DNP dynamics are caused by the inverse squared relationship between the triple spin flip rate $W_{\text{SSI}}^{\pm}$ and $\omega_{\text{I}}$ in equation 2, being 42.57 MHz at 1 T for $^1$H.

## 2.3 DNP simulations

Finally, the aim is to predict the CE and SE DNP spectra by estimation of the DNP induced enhancement of $^1$H and $^{13}$C spins by the extent of first- and second-order saturation of electrons. We employed a numerical model that calculates reduction of polarization (saturation) of $N$ discretised electron bins under microwave irradiation. The time-dependent master equation for electron saturation under microwave irradiation is shown in equation 4 and is inspired by the methodology implemented by Bloch[34], Torrey[35], Hovav et al.[23,36,37] and finally Wenckebach[38] to model the depolarization dynamics of the electron. Here, the electron saturation model considers the influence of:

(i) the microwave irradiation position ($\omega_{\text{m}}$),
(ii) the microwave field strength ($\omega_{1S}$),
(iii) the electron spin-lattice relaxation constant ($T_{1S}$),
(iv) the electron spin-spin relaxation constant ($T_{2S}$) and,
(iv) the electron spectral diffusion constant ($\mathfrak{D}$),

on the polarization of each electron bin over time $P_S(\omega_S, t)$. The spectral diffusion constant $\mathfrak{D}$ represents the rate of indirect spread of non-equilibrium polarization in between neighbouring electron bins by dipolar-induced flip-flop transitions.[37–39] Polarization transfer from electrons to the nuclear spins is then estimated based on specific matching conditions in between the depolarized electron bins constituting the EPR line.[40] Both selection criteria are summarized in equations 9 and 11 for SE and CE, and are discussed *vide infra*. The equations were adapted from the electron bin model from Hovav et al.[23] that has been proven a robust way to model solid-state DNP at amorphous conditions with high concentrations of radicals.[37,39,41] The codes used for the simulation of the DNP spectra are made available in the data repository[42] and are written in MATLAB (The MathWorks, Inc, Massachusetts, USA).

Table 1. g-tensor and hyperfine (A) tensor (with $^{14}$N) values for TEMPOL and trityl OX063. The values for TEMPOL were extracted from the experimental CW-EPR spectrum on a 50 mM TEMPOL sample as shown previously[13] and again in Figure S1. For trityl OX063, g-tensor values were taken from the literature with a Gaussian and Lorentzian broadening of respectively 5 and 2 MHz.[43]

| TEMPOL | | | | Trityl OX063 | | | |
|---|---|---|---|---|---|---|---|
| g-tensor (-) | | A-tensor (MHz) | | g-tensor (-) | | A-tensor (MHz) | |
| $g_{xx}$ | 2.008 | $A_{xx}$ | 24.23 | $g_{xx}$ | 2.0032 | $A_{xx}$ | / |
| $g_{yy}$ | 2.005 | $A_{yy}$ | 16.96 | $g_{yy}$ | 2.0032 | $A_{yy}$ | / |
| $g_{zz}$ | 2.001 | $A_{zz}$ | 103.33 | $g_{zz}$ | 2.0026 | $A_{zz}$ | / |

### 2.3.1 Electron spins

Prediction of the $^1$H and $^{13}$C DNP spectra starts with finding the EPR lineshape of the radicals under DNP conditions. For TEMPOL, the g-tensor and hyperfine A-tensor of an identical 50 mM TEMPOL DNP sample were extracted by fitting a CW-EPR spectrum using the MATLAB-based package Easyspin[44] measured at very close conditions of 1.2 T (Q-band, 34 GHz) and 77 K as reported before[13] and shown again in Figure S1. For trityl OX063, g-tensor values were taken from the literature.[43] For both radicals, the values are reported in Table 1. In Figure 4a, the simulated EPR profile of trityl OX063 is shown in black. The symmetric chemical structure of the trityl OX063 radical combined with the absence of any strong hyperfine coupling with other nuclear spins in the vicinity of the unpaired electron results in the narrow EPR lineshape positioned at $\omega_S$ of 28.05 GHz in benchtop DNP conditions. In Figure 4b, the EPR profile of TEMPOL is shown and is much broader as compared to trityl OX063 (notice the frequency scale difference) due to the large anisotropy of the g-tensor and the field-independent strong hyperfine interaction due to the presence of $^{14}$N close to the unpaired electron. For both radicals, in grey, the electron bins ($N$ = 128) with a homogeneous Lorentzian line shape $h$ in equation 3 are plotted underneath:

$$h(\omega_S - \omega) = \frac{1}{\pi} \frac{T_{2S}}{1 + (\omega_S - \omega)^2 T_{2S}^2}. \quad (3)$$

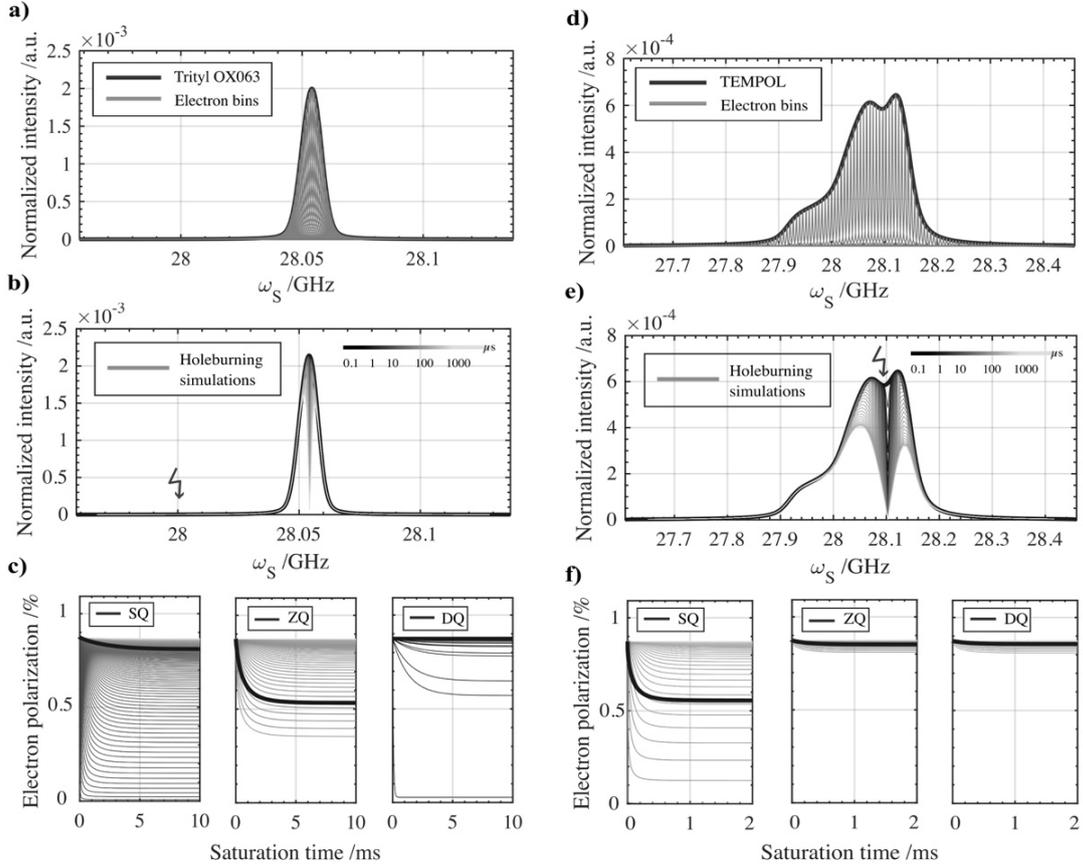

**Figure 4. Time-resolved electron saturation simulations during microwave irradiation for DNP.** The master equation 4 for electron saturation dynamics under microwave irradiation is numerically solved and considers the influence of the microwave position ($\omega_m$), the irradiation strength ($\omega_{1S}$), the electron spin-lattice ($T_{1S}$) and spin-spin relaxation ($T_{2S}$), and spectral diffusion ($\mathfrak{D}$) on the polarization of discretised electrons bins over time $P_S(\omega_S,t)$. The simulations parameters are summarized in Table 2. **a,d)** Discretized solid-state EPR spectrum of trityl OX063 (left) and TEMPOL (right). The EPR line shape was simulated using the MATLAB-based Easyspin package. For TEMPOL, the anisotropic g-tensor and hyperfine A-tensor of an identical 50 mM TEMPOL DNP sample were extracted by fitting a CW-EPR. For trityl OX063, g-tensor values were taken from literature and a Gaussian/Lorentzian homogenous broadening of 5/2 MHz was added. All tensor parameters used to fit the EPR spectra are shown in Table 1. The areas underneath the EPR line were both normalized to 1. **b,e)** Microwave hole burning simulations for trityl OX063 (left) at $\omega_m$ of 28.0 GHz and TEMPOL (right) at $\omega_m$ of 28.1 GHz during 10 ms with $\omega_{1S}/2\pi$ of 500 kHz. **c,f)** Electron polarization of the electron bins evolving over time. The grey lines show the electron polarization of the separate bins $P_S(\omega_S,t)$ over time, while the black lines show the frequency-weighted electron polarization average $P_S^{av}$ plotted separately for the contribution of SQ, ZQ and DQ transitions. Result from the electron saturation simulation can be reproduced by the MATLAB codes made available at the repository.[42]

To describe the dynamics of electron saturation under microwave irradiation, the master rate equation 4 was numerically solved (codes available[42]) by employing the ordinary differential equation (ODE15s) solver for stiff problems in MATLAB.

$$\frac{\partial P_S(\omega_S,t)}{\partial t} = \mathfrak{W} P_S(\omega_S,t) + \mathfrak{D}\frac{\partial^2 P_S(\omega_S,t)}{\partial \omega_S^2} + \mathfrak{R}\big(P_L - P_S(\omega_S,t)\big). \quad (4)$$

This allowed to calculate the electron polarization $P_S(\omega_S,t)$ at time $t$ for electrons at Zeeman frequency $\omega_S$ after microwave saturation at frequency $\omega_m$. The first term in equation 4 represents the off-resonant microwave saturation rate[35,38] with $\mathfrak{W} = \mathfrak{W}_{SQ} + \mathfrak{W}_{ZQ,DQ}$ including first-order (single quantum) electron transition rates $\mathfrak{W}_{SQ}$:

$$\mathfrak{W}_{SQ} = \omega_{1S}^2 h(\omega_s - \omega_m) = \omega_{1S}^2 \frac{T_{2S}}{1+(\omega_S-\omega_m)^2 T_{2S}^2}, \quad (5)$$

and second-order (zero-quantum, double-quantum), nuclear electron transition rates $\mathfrak{W}_{ZQ,DQ}$:

$$\begin{aligned}\mathfrak{W}_{ZQ,DQ} &= \pi\left(\frac{A_{IS}}{2\omega_I}\omega_{1S}\right)^2 h(\omega_S \pm \omega_I - \omega_m) \\ &= \left(\frac{A_{IS}}{2\omega_I}\omega_{1S}\right)^2 \frac{T_{2S}}{1+(\omega_S \pm \omega_m)^2 T_{2S}^2}.\end{aligned} \quad (6)$$

The second term describes the change of electron polarization caused by electron flip-flop transitions when a polarization gradient over $\omega_S$ is present. This transport process can be modelled as a random walk and therefore $\mathfrak{D}$, the spectral diffusion constant, can be regarded similar to the thermal diffusivity constant during one-dimensional heat transport.[38] The last term represents the spontaneous evolution towards Boltzmann equilibrium polarization $P_L$ described by the spin-lattice electron relaxation rate constant:

$$\Re = \frac{1}{T_{1S}}. \tag{7}$$

Depolarization modelling was performed by estimating the rate constants for saturation, spectral diffusion and spin-lattice relaxation in equation 4 using the electron and nuclear spin properties of TEMPOL and trityl OX063 at benchtop conditions summarized in Table 2. For TEMPOL, the spin-lattice electron relaxation time $T_{1S}$ of the nitroxide radicals was estimated by pulsed EPR saturation recovery experiments measured at Q-band in Figure 6d. The results are shown in Figure S2 and Figure S3. The empirically observed relaxation time constants were extracted after being fitted with the stretched exponential buildup from equation 15. $T_{1S}$ ranges from 195 µs to 423 µs and is anisotropic with respect to the electron frequency as previously observed.[36,45,46] The spin-spin electron relaxation time $T_{2S}$ was estimated using the Carr-Purcell-Meiboom-Gill (CPMG) pulse train in Figure 6e and fitted with the stretched exponential in equation 17. $T_{2S}$ ranges from 0.75 µs to 0.87 µs for the field positions shown in Figure S3 and even faster $T_{2S}$ are expected at the field positions where no echo intensity could be measured after 0.8 µs. For the simulations, only the average of the characteristic electron relaxation time was taken. Again, for trityl OX063, characteristic electron relaxation values were estimated based on the literature.[23,24,43,47]

**Table 2.** Electron and nuclear spin constants required to calculate the saturation of electrons under microwave irradiation for TEMPOL and trityl OX063. $\mathfrak{D}$ was initially estimated based on a recent prediction of the spectral diffusion constant for 40 mM TEMPOL by Wenckebach[38] at 3.34 T and 20 K. For trityl OX063, $\mathfrak{D}$ was taken to fit the experimental ¹H DNP spectrum. $A_{IS}$ values (with ¹H) were based on values reported in literature [23,24,43,47] and for ¹³C were taken 1 order of magnitude lower. The microwave field strength $\omega_{1S}$ was estimated by using CST Microwave Studio simulations reported in Bocquelet et al.[11]

| Constants | TEMPOL | Trityl OX063 | Units |
|---|---|---|---|
| $\omega_I$ | 2π·42.58·10⁶ | 2π·42.58·10⁶ | rad·s⁻¹ |
| $T_{1S}$ | 0.309·10⁻³ | 2.00·10⁻³ | s |
| $T_{2S}$ | 0.81·10⁻⁶ | 1.00·10⁻⁶ | s |
| $\mathfrak{D}$ | 170·10¹⁸ | 1·10¹⁸ | s⁻³ |
| $A_{IS}$ | 2π·2·10⁶ | 2π·1·10⁶ | rad·s⁻¹ |
| $P_L$ | 0.0088 | 0.0088 | / |
| $\omega_{1S}$ | 2π·500·10³ | 2π·500·10³ | rad·s⁻¹ |

The results of the electron saturation simulations for TEMPOL and trityl OX063 are shown in Figure 4. The evolution of the electron saturation is simulated during continuous microwave irradiation at 28.1 GHz in case of TEMPOL and at 28.0 GHz in case of trityl OX063. The change in electron polarization $P_S(\omega_S, t)$ is plotted over time for each electron bin in Figure 4c for trityl OX063 (2 ms) and Figure 4f for TEMPOL (10 ms). The solid black curve illustrates the average polarization $P_S^{av}$ after microwave irradiation and is calculated by taking the weighted average over all the electron bins. $P_S^{av}$ is already a good indicator for which DNP mechanism can be efficient. Immediately noticeable is that under identical conditions saturation by second-order transitions is much more efficient for trityl OX063 ($\Delta P_S^{av} = -0.37$ %) than in case of TEMPOL ($\Delta P_S^{av} = -0.01\%$). This is mainly due to the narrower line shape of trityl OX063 compared to TEMPOL making it easier for the microwave to depolarize many electrons at once. Moreover, due to the narrow line shape of trityl OX063, second-order transitions can be excited independently. SE DNP that relies on driving these electron-nuclear second-order transitions can therefore produce positive and negative polarization at corresponding microwave positions without cancelling out.

In contrast, for TEMPOL, the broad EPR line results in simultaneous excitation of the SQ, ZQ and DQ transitions along most of the EPR line. This is illustrated in Figure 4e, where the microwave initially burns a narrow hole at 28.1 GHz which broadens rapidly within a few microseconds as the process of spectral diffusion spreads the polarization difference along the EPR line. As observed, the polarization gradient rapidly reaches steady state on a millisecond timescale much shorter than the typical CW microwave irradiation time used in the experiments (seconds). In certain cases (see materials and methods section), the steady-state solution of equation 4 can therefore be estimated analytically by:

$$P_S(\omega_s) = P_L \left[1 - s \cdot \exp\left(-\frac{|\omega_S - \omega_m|}{\sqrt{\mathfrak{D} T_{1S}}}\right)\right], \tag{8}$$

where $s$ is a relative saturation factor.[48] In Figure S4, both solutions obtained either via numerical or analytical calculations agree for 50 mM TEMPOL. By looking at the analytical expression in equation 8, the parameters influencing the extent of electron saturation and therefore indirectly also the DNP efficiency become clear. The width of the hole is dictated by the $\sqrt{\mathfrak{D} T_{1S}}$ term in denominator of the exponential function. Fast spectral diffusion $\mathfrak{D}$ in combination with a long electron spin-lattice relaxation time $T_{1S}$ are therefore ideal for increasing the overall electron saturation. The radical concentration, the radical type and the lattice temperature all influence $\sqrt{\mathfrak{D} T_{1S}}$ and therefore are key parameters for optimal saturation efficiency.[13,22,43,49] At our benchtop conditions, the short $T_{1S}$ of 300 µs (Table 2) thus acts as a limiting factor for driving first-order transitions. Second-order transitions are present as well for TEMPOL as shown in Figure 3f, but are even less efficiently driven ($\Delta P_S^{av} = -0.01$ %) and most importantly are equally driven for ZQ and DQ transition. While this is detrimental for SE DNP, the reduced polarization of these electron spins can still lead to CE type of DNP when the requirements of triple spin flips with other neighboring electrons are met: hyperpolarization of nuclear spins can only happen if the energy needed to flip and thus polarize a nuclear spin ($\omega_I$) can be compensated by a simultaneous flip-flop of the electron pair.[19,50] The extent of polarization that can be transferred to nuclear spins is presented next, which predicts and rationalizes the ¹H and ¹³C DNP spectra.

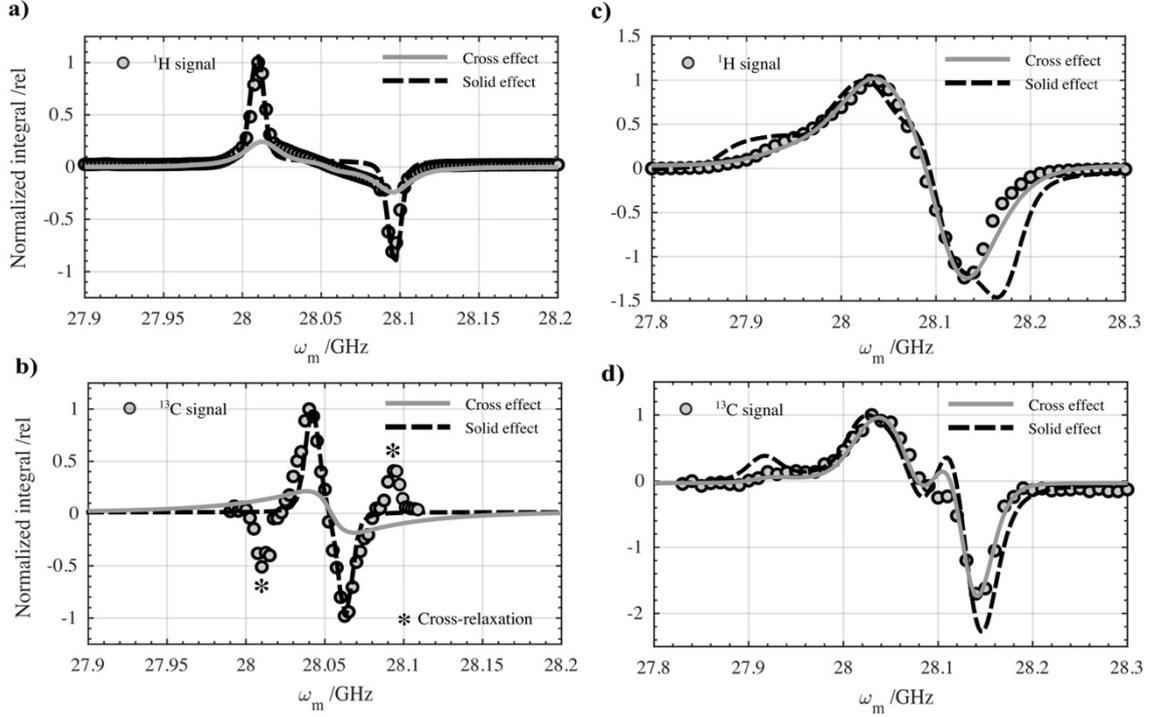

**Figure 5. Normalized experimental and simulated SE and CE DNP spectra at 1 T and 77 K for proton ($^1$H) and carbon ($^{13}$C) nuclear spins based on the electron saturation simulations.** The normalized nuclear spin polarization generated at each frequency were analytically calculated based on the SE and CE section criteria for the electron bins in respectively equations 9 and 11. **a,c)** The calculated $^1$H DNP spectrum for trityl OX063 (left) and TEMPOL (right) assuming SE and CE polarization transfer overlayed with the experimental $^1$H DNP spectrum as a comparison. **b,d)** The calculated $^{13}$C DNP spectrum for trityl OX063 (left) and TEMPOL (right) assuming SE and CE polarization transfer overlayed with the experimental $^{13}$C DNP as a comparison. The additional DNP optima indicated with an asterisk could not be reproduced by the DNP simulations but were assigned to spontaneous $^1$H-$^{13}$C heteronuclear cross relaxation in acetate. Codes used for the DNP spectra simulations are made available[42] and are written in MATLAB.

### 2.3.2 Nuclear spins

The average nuclear spin polarization $P_I(\omega_m)$ produced by irradiating at $\omega_m$ can be calculated considering the extent of saturation by first- and second-order transitions and the selection criteria for either SE or CE polarization transfer shown in equations 9 and 11 respectively. For SE DNP, the extent of second-order transitions happening at electron bins around $\omega_s = \omega_m \pm \omega_I$ allows to predict $P_I(\omega_m)$ by:

$$P_I(\omega_m) = \frac{1}{f} \sum_{j=1+\Delta j}^{N-\Delta j} \left[ p_{j+\Delta j} \frac{P_{S,j+\Delta j} - P_{S,j}}{1 - P_{S,j} P_{S,j+\Delta j}} - p_{j-\Delta j} \frac{P_{S,j-\Delta j} - P_{S,j}}{1 - P_{S,j} P_{S,j-\Delta j}} \right], \quad (9)$$

with $p_j$ and $P_{S,j}$ the respective fraction and polarization of electrons at bin $j$. The shift $\Delta j$ represents the number of electrons bins with a frequency difference of $\omega_I$ at which second-order transitions occur and $f$ is a normalization factor:

$$f = \sum_{j=1+\Delta j}^{N-\Delta j} p_{j+\Delta j}. \quad (10)$$

Positive nuclear spin polarization produced by DQ transitions ($\omega_s = \omega_m - \omega_I$) are then summed together with the negative nuclear spin polarization produced by ZQ transitions ($\omega_s = \omega_m + \omega_I$) at each electron bin along the EPR line. Note that here we don't consider the contribution of SQ transitions on the saturation of electrons. $P_{S,j}$ is therefore equal to the Boltzmann electron polarization.

For CE, DNP depends on both first- and second-order transition to generate a polarization imbalance in between neighboring electron pairs that satisfies the triple spin flip matching condition. While the triple spin flip matching condition can be satisfied via different kinds of interaction energy, such as dipolar energy, exchange energy, or Zeeman energy,[32,50] this model uses solely the Zeeman energy difference between electrons, the dominant interaction energy for monoradicals, to simulate the hyperpolarization of nuclear spins.[19,31] The average nuclear spin polarization $P_I(\omega_m)$ after irradiation at $\omega_m$ can then be estimated by:

$$P_I(\omega_m) = \frac{1}{f} \sum_{j=1+\Delta j}^{N-\Delta j} \left[ p_j p_{j+\Delta j} \frac{P_{S,j+\Delta j} - P_{S,j}}{1 - P_{S,j} P_{S,j+\Delta j}} - p_j p_{j-\Delta j} \frac{P_{S,j-\Delta j} - P_{S,j}}{1 - P_{S,j} P_{S,j-\Delta j}} \right], \quad (11)$$

with $f$ again the normalization factor:

$$f = \sum_{j=1+\Delta j}^{N-\Delta j} p_j p_{j+\Delta j}. \quad (12)$$

This method is equivalent to the indirect CE model proposed by Hovav *et al.*[41] and calculates the polarization imbalance of the fraction of electrons at $p_j$ and $p_{j\pm\Delta j}$ for producing both positive and negative nuclear spin polarization. While the value of $\Delta j$ remains identical as with calculating SE DNP, here it represents the triple spin flip matching criteria, thus the shift in number of bins that is required to match the frequency difference of the electron pair to the Larmor frequency $\omega_I$ of the nuclear spin under observation.

The results for the $^1$H and $^{13}$C DNP spectra simulations compared to experimental data are presented in Figure 5. For TEMPOL, the shapes of the experimental $^1$H and $^{13}$C DNP spectra in Figure 5c and Figure 5d agree well with the predicted CE DNP spectra (solid line) in contrast to the predicted SE DNP spectra (dotted line). This fortifies the statement that for both $^1$H and $^{13}$C nuclei CE is dominant for 50 mM nitroxide concentration at 1 T and 77 K. To allow for the best fit, the spectral diffusion constant was fitted to match the experimental $^1$H DNP spectrum using a least square refinement. The resulting $\mathfrak{D}_{\text{fitted}}$ was $78 \cdot 10^{18}$ s$^{-3}$ and is similar to what is predicted by Wenckebach[38] and on the same order of magnitude as what was experimentally derived before by Kundu *et al.*[37] for 40 mM TEMPOL, as summarized in Table 3.

**Table 3.** Experimentally fitted spectral diffusion constant $\mathfrak{D}$ for 50 mM TEMPOL at 1 T and 77 K. The value of $\mathfrak{D}$ is compared against the recently theoretically predicted $\mathfrak{D}$ for 40 mM TEMPOL by Wenckebach[38] and the experimentally derived $\mathfrak{D}$ for 40 mM TEMPOL by Hovav *et al.*[23]

|  | This work | Wenckebach[38] | Hovav *et al.*[23] |
|---|---|---|---|
| $\mathfrak{D}$ (s$^{-3}$) | $78 \cdot 10^{18}$ | $340 \cdot 10^{18}$ | $170 \cdot 10^{18}$ |
| [TEMPOL] | 50 mM | 40 mM | 40 mM |

Surprisingly also for trityl OX063 in Figure 5a, besides the reproduced narrow SE DNP maxima, CE polarization transfer seems to play a role at 1 T and 77 K and fits well with the appearance of the broad component in the $^1$H DNP spectrum of Figure 5a. The reason could be that due to the lower magnetic field, electron trityl pairs matching the proton Larmor frequency difference (42.57 MHz) are found more easily.[47] In the saturation simulations presented in Figure 4c, off-resonant microwave irradiation in combination with spectral diffusion is indeed able to depolarize a fraction of the electron spins in the case of trityl OX063, thus potentially leading to CE DNP. When following the same reasoning for hyperpolarization of $^{13}$C as seen in Figure 4b, CE matching conditions for neighboring electron pairs should be even less selective, and therefore boost the efficiency of CE DNP for trityl OX063 even more. However, due to the complexity of the experimental results of the $^{13}$C DNP, the contribution of CE DNP in $^{13}$C even with the help of the simulations is difficult to verify in trend with reports from Michaelis at 5 T.[51] First, it seems that SE DNP reproduces well the main dispersive feature at 28.05 GHz. This is in contradiction with Banerjee and co-workers that predicted the presence of CE in $^{13}$C nuclei for 15 mM and 30 mM trityl OX063 even at a higher magnetic field of 3.5 T.[47] On top, instead of the underlying broader contribution predicted for CE DNP, additional optima (indicated with an asterisk) with opposite polarization to the dispersive peak in the middle are observed in the $^{13}$C DNP spectrum and obscure the potential contribution of CE DNP.

A closer investigation showed that efficient heteronuclear cross relaxation between dipolar coupled $^1$H and $^{13}$C spins is most probably responsible for their contribution, as pointed out recently as well at 8.9 T and 100 K using trityl OX063 by Palani *et al.*[30] Other SE-based DNP mechanisms that can introduce additional DNP maxima such as multiple-spin solid effect[29] and $^1$H-$^{13}$C heteronuclear CE[52] appearing in a four-spin system are disregarded by the following reasons. First, the position of the additional $^{13}$C DNP optima are situated only at $4 \cdot \omega_I^{13C}$ and no multiples of $n \cdot \omega_I^{13C}$ with n = 1, 2, 3 as would be the case in the multiple-spin solid effect scenario. Their appearance corresponds with the microwave position where proton DNP is optimal. This hints that the additional $^{13}$C magnetization initially originates from proton nuclear spins. Saturation of protons during $^{13}$C acquisition at the optimum DNP frequency at 28.0075 GHz indeed removed the carbon signal in line with our proposition as shown in Figure S5. Secondly, heteronuclear dipolar induced cross relaxation dominated by double quantum relaxation can explain why carbon polarization for the additional extrema is opposite in sign from that of the protons.[53] This process is equivalent to nuclear Overhauser effect (NOE) type of polarization transfer[54], yet instead of creating the polarization imbalance between nuclear spins by means of saturation[55], hyperpolarization of protons is responsible for it and the imbalance in zero- and double-quantum relaxation gives rise to polarization transfer. In contrast, the $^1$H-$^{13}$C heteronuclear CE would result in a dispersive line shape at this microwave position.[52] Fast methyl rotation in the μs timescale is present at cryogenic temperatures and is known to have rotational frequencies in the order of MHz that can lead to efficient double quantum relaxation.[56–58] In magic angle spinning (MAS) DNP at higher fields, this has been shown to induce spontaneous heteronuclear cross relaxation, and has even been employed as a solid-state structural elucidation methodology since the effect is distance-dependent.[27,59,60] Due to the strong contribution, the current benchtop DNP polarizer could therefore help in further investigating this spontaneous polarization transfer phenomenon between different nuclei.

## 3. Conclusions

In conclusion, we conducted a DNP study searching for the dominant polarization transfer mechanism for narrow- and broad-line DNP polarizing agents at 1 T and 77 K. The lower field and higher temperature compared to conventional dDNP are required for developing a benchtop DNP device capable of hyperpolarized flow DNP.[11] The impact of the microwave power on the $^1$H DNP buildup times and $^1$H spin-lattice relaxation times were important for deducing the dominant DNP mechanisms. Electron saturation simulations were developed to calculate first- and second-order saturation efficiency of trityl OX063 and TEMPOL at 1T and 77 K. The DNP simulation results assisted in analysis of the $^1$H and $^{13}$C DNP spectra. We found out that for 50 mM TEMPOL, a representative of broad-line nitroxide radical family, cross effect is the dominant polarization mechanism both for $^1$H and $^{13}$C. We observed that $^1$H DNP buildup times equal the nuclear spin relaxation times independently of the microwave power. This observation points towards triple spin flips being the rate limiting step in the DNP process even at 1 T and 77 K, and dominating both hyperpolarization and nuclear spin-lattice relaxation. Finally, electron saturation simulations starting from experimentally measured EPR lineshapes and spin-lattice and spin-spin relaxation times of TEMPOL at Q-band accurately matched the measured $^1$H and $^{13}$C DNP spectra in case of cross effect.

For 15mM trityl OX063, a representative narrow-line radical, DNP at 1 T and 77 K is mostly dominated by the solid effect, but both in $^1$H and $^{13}$C other routes of polarization transfer are noticed as well. Overall, hyperpolarization is clearly produced at zero- and double-quantum electron-nuclear transitions as seen from the position of the DNP optima in the $^1$H and $^{13}$C DNP spectra. Moreover, increasing the microwave power reduces the DNP buildup time in line with the microwave-driven solid effect hyperpolarization process. However, for $^1$H, cross effect DNP seems to be active as well and is responsible for a broad component in the $^1$H DNP spectrum. This was reproduced by saturation simulations that did show the ability for first-order saturation far away from the central EPR line (> 40 MHz) to depolarize trityl OX063 electrons. The lower magnetic field of 1 T is thought to be responsible for making it easier to find electron pairs with the correct matching conditions for cross effect. Finally, in case of $^{13}$C, heteronuclear cross relaxation between dipolar coupled $^1$H and $^{13}$C spins in [1-$^{13}$C] acetate seems effective at 1 T and 77 K and cause the appearance of additional DNP optima at the microwave position where proton DNP is optimal.

In the future, we plan to devise strategies to boost sensitivity even further having knowledge on the dominant DNP mechanisms and on the parameters that restrict us (microwave power and short electron spin-lattice relaxation time). We would also like to expand the DNP model to be able to predict absolute polarization levels, and to theoretically study the influence of varying electron relaxation properties, the spectral diffusion efficiency, and the radical type (biradicals) at benchtop conditions.

## 4. Materials and methods

### 4.1 Solid-state DNP experiments

#### 4.1.1 Amorphous sample formulation for trityl OX063 and TEMPOL frozen samples

A partially protonated DNP solution with the following volumetric ratio 2:2:6 H$_2$O:D$_2$O:DMSO-d$_6$ ($v$:$v$:$v$) was doped with either 15 mM trityl OX063 or 50 mM TEMPOL (4-hydroxy-2,2,6,6-tetramethylpiperidin-1-oxyl) to allow efficient glass formation upon freezing. The employed 50 mM radical concentration for TEMPOL corresponds to the optimal concentration giving the largest enhancement factor in HypFlow DNP conditions[13], while for trityl OX063, the optimal concentration of 15 mM was chosen based on previous reports.[61] All chemicals other than the trityl OX063 radical were purchased from Sigma-Aldrich. The deuteration purity for D$_2$O and DMSO-d$_6$ were respectively 99.5% and 99.96%. 200 μL of the DNP solution was filled in a 4 mm (OD) Wilmad quartz EPR tube and placed in the benchtop DNP polarizer. The final amount of proton spins eligible for hyperpolarization amounts to 2.2 mmol or a concentration of 22 M in both DNP samples.

The $^{13}$C DNP samples were prepared analogously to the samples used for $^1$H DNP starting from the same 2:2:6 H$_2$O:D$_2$O:DMSO-d$_6$ ($v$:$v$:$v$) glass-forming solution. Additionally, a total of 3 M labelled [1-$^{13}$C] sodium acetate was added and ultimately the $^{13}$C DNP samples were doped with either 15 mM trityl OX063 or 50 mM TEMPOL.

#### 4.1.2 Benchtop DNP polarizer

Hyperpolarization experiments were performed using a prototype benchtop DNP polarizer co-developed with Bruker Biospin (Figure 1). The DNP polarizer is equipped with (i) a temperature-regulated permanent 1 T magnet, (ii) a cryostat insert operating at 77 K using liquid nitrogen, (iii) a home-built $^1$H/$^{13}$C NMR probe composed of a saddle coil with orthogonal $B_1$ fields of 2.9 kHz/W$^{\frac{1}{2}}$ for proton and a solenoid coil with 3.9 kHz/W$^{\frac{1}{2}}$ for carbon resonating at the frequencies of respectively 42.57 MHz and 10.7 MHz, and (iv) a solid-state Kα-band microwave source that amplifies microwave frequencies from 26.6 to 28.8 GHz up to 5 ± 2 watts. Once generated and amplified, a WR-28 coax-to-waveguide transition excites the TE$_{01}$ rectangular mode, and through a standard gain microwave horn the microwave is irradiated on the frozen sample. More information on the design and specifications of the setup can be found in Bocquelet et al.[11]

#### 4.1.3 Experimental design

The $^1$H and $^{13}$C DNP spectra in Figure 2 were constructed based on the pulse sequence shown in Figure 6a and 6b. For $^1$H, a hard $\pi/2$ rf pulse was used to maximize the detected hyperpolarized signal and for $^{13}$C, detection happened via a solid echo using a 14 μs echo time between the middle of the $\pi/2$ rf pulses. After acquisition, a TTL pulse generated by the NMR console was able to trigger the microwave source and change frequency by 10 MHz within a given delay (5 s for TEMPOL and 60 s for trityl OX063). The microwave frequency was swept from 27.8 to 28.44 GHz for all samples, except for trityl OX063 with sodium acetate added where the microwave frequency was swept from 29.975 to 28.115 GHz. The DNP spectra were constructed using MATLAB by plotting the integration values of the hyperpolarized signals against the varying microwave frequency ($\omega_\mathrm{m}$). Since the MATLAB processing pipeline only imported frequency domain NMR data, the fast Fourier transform, and line broadening were performed using Topspin.

The characteristic hyperpolarization buildup times for TEMPOL and trityl OX063 samples were extracted using the pulse sequence in Figure 6c. First, residual magnetization on the $^1$H rf channel was destroyed by a saturation block using hard $\pi/2$ rf pulses with alternating phases that are separated by a short delay and repeated 50 times. The hyperpolarization buildup was the monitored via small $\pi/36$ flip-angle rf pulse that allowed to trace the $^1$H NMR signal when the microwave was switched on at variable power of 0.55, 1.32, 2.3, 3.16, 3.8, 4.27, 4.57 W using a potentiometer. A total of 25 small angle rf pulse were used to monitor the buildup with a delay τ of 0.3 s in the case of TEMPOL and 1 s in the case of trityl OX063. After buildup, the microwave source was switched off and the decay was monitored identically. All processing of the time-domain data was performed identically using 8192 datapoints in the F2 dimension and the same exponential line broadening of 1 kHz. Next, the extraction of the DNP buildup times $\tau_{\text{DNP}}$ and nuclear spin relaxation times $T_1$ was performed using MATLAB by fitting the integrated small flip-angle signals ($I_t^{\text{DNP}}$) according following exponential buildup and decay:

$$I_t^{\text{DNP}} = I_\infty^{\text{DNP}} \left(1 - \exp\left(-\frac{t}{\tau_{\text{DNP}}}\right)\right), \quad (13)$$

$$I_t = I_0^{\text{DNP}} \left(\exp\left(-\frac{t}{T_1}\right)\right). \quad (14)$$

The error margins depicted in Figure 3 correspond to 95% certainty interval.

### 4.2 Q-band EPR experiments at 1.2 T and 77 K

#### 4.2.1 Sample formulation

The continuous wave (CW) and pulsed EPR experiments were carried out with TEMPOL in frozen DNP solution, identical to solutions used in DNP benchmarking experiments and capable of glass formation. TEMPOL (4-hydroxy-2,2,6,6-tetramethylpiperidin-1-oxyl), ≥97% total nitrogen (N) basis, and DMSO-$d_6$ (hexadeuterodimethyl sulfoxide, 99%) were purchased from Sigma-Aldrich, Merck KgaA, Darmstadt, Germany. A stock solution of 100 mM TEMPOL was first prepared by dissolving 3.516 mg of TEMPOL into 204.1 µL into the glass-forming solute with a volumetric ratio of 2:2:6 H$_2$O:D$_2$O:DMSO-$d_6$ (v:v:v). After dilution to the desired radical concentration (50 mM), 1 µL of the DNP solution was taken in a 1.6 mm (o.d.) quartz tube to perform the EPR experiments.

#### 4.2.2 Instrumentation

EPR experiments were performed at 77 K and 34 GHz/1.2 T/ 51 MHz on a Bruker Elexsys E580 spectrometer with Q-band extension (SuperQ-FT-u bridge). The EN 5107D2 EPR/ENDOR probe with a dielectric resonator capable of performing both CW and pulse EPR experiments was used. The pulse programmer was a PatternJet II with 2 ns resolution and the pulses were amplified with a BLA50 ESR 33–35 GHz 50 W amplifier (Bruker BioSpin GmbH). In both cases, the sample temperature was kept at 77 K using a CF 935 flow cryostat (Oxford Instruments) and liquid helium as a cryogen.

#### 4.2.3 Experimental design

**Continuous-wave (CW) and echo detected pulse EPR**
Field-swept CW EPR experiments shown in Figure S1 were conducted with a field modulation amplitude of 0.5 G at a modulation frequency of 50 kHz for phase-sensitive detection of the reflected microwave signal. The microwave power was set to 60 µW (25 dB attenuation), and the central microwave frequency was 34.1842 GHz. The echo-detected pulse EPR profile on a 50 mM TEMPOL frozen DNP solution in Figure S2c was measured with a central microwave frequency of 34.03562 GHz. More information on the pulse sequence used can be found in Figure S2b in the supporting information.

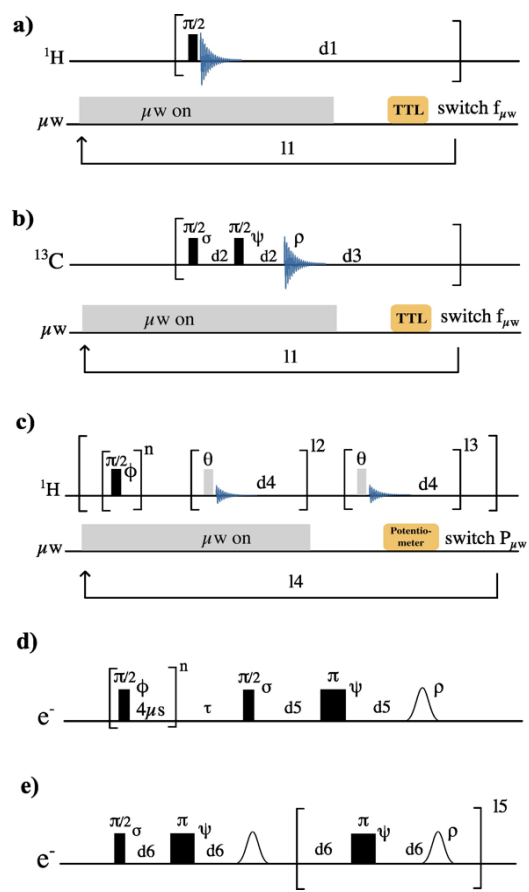

**Figure 6.** Pulse sequences employed to monitor nuclear and electron spin dynamics in the frozen radical solution. **a-b)** Pulse sequence to monitor either the $^1$H DNP spectrum (a) or the $^{13}$C DNP spectrum (b) by changing the microwave frequency by 10 MHz for TEMPOL and 2.5 MHz for trityl OX063. The $^1$H DNP spectrum is monitored using a single π/2 pulse with d1 = 5 s and l1 = 65 for TEMPOL and d1 = 60 s and l1= 160 for trityl. The $^{13}$C spectrum is monitored using a solid echo with d2 = 14 µs in between the center of the π/2 pulses and d3 = 60 s with an 8-step phase cycling with σ = +x, -y, -x, y, +x, -y, -x, y and ψ = -y, -x, y, x, y, x, -y, -x and with detection phase ρ = -y, -x, y, x, -y, -x, y, x. **c)** Pulse sequence to monitor the DNP buildup rate ($\tau_{\text{DNP}}$) and nuclear spin relaxation time ($T_1$) at variable microwave power (0.55, 1.32, 2.3, 3.16, 3.8, 4.27, 4.57 W) using small-angle θ rf pulses of π/36. For saturation a 4-step phase cycling is used with φ = +x, +y, -x, -y and n = 50. For TEMPOL, d4 = 0.3 s, l2 = 25, l3 = 25, l4 = 7. For trityl OX063, d2 = 1 s, l2 = 25, l3 = 25, l4 = 7. **d)** Saturation recovery pulse sequence to extract the electron spin-lattice relaxation constants ($T_{1S}$). For saturation a 4-step phase cycling is used with φ = +x, +x, -x, -x and n = 20. Afterwards, the variable delay is τ and echo detection is performed with d5 = 800 ns. Phase cycling is σ = +x, -x, +x, -x and ψ = +x, +x, +x, +x with echo detection phase ρ = +, -, +, -. **e)** Pulse sequence to extract the electron spin-spin relaxation constants ($T_{2S}$) using a CPMG sequence with a two-step phase cycle σ = +x, -x and ψ = +y, +y with echo detection phase ρ = +, -. Here d6 = 390 ns and l5 is 16.

**Electron spin-lattice and spin-spin relaxation**
Spin-lattice electron relaxation times were measured at two different regions of the EPR spectrum using a saturation recovery pulse train (Figure 6d) with a variable delay between the saturation block (series of 90° pulses) and the echo-detection scheme $(t_{90})_x - \tau - (t_{180})_x - \tau$ – detection. A $t_{90}$ of 14 ns was chosen with an echo delay τ of 800 ns. The resulting electron spin magnetization buildup curve was then constructed by integration of EPR echo intensity and fitted with a minimal least-square error using a stretched exponential:

$$I_t^{\text{EPR}} = I_\infty^{\text{EPR}} \left(1 - \exp\left(-\frac{t}{T_{1S}}\right)^\beta\right), \quad (15)$$

where $I_\infty^{\text{EPR}}$, $T_{1S}$ and $\beta$ are respectively the EPR signal intensity at infinite relaxation time, the empirical spin-lattice electron relaxation time constant, and the stretch coefficient between 0 and 1. The stretch coefficient is an indication of whether a single ($\beta = 1$) or multiple ($\beta < 1$) dominant mechanisms are causing spin-lattice relaxation of the electrons. In the latter case, a distribution of relaxation time constants is expected for which the average relaxation time constant can be calculated using:

$$T_{1S,\text{av}} = \frac{T_{1S}}{\beta} \Gamma\left(\frac{1}{\beta}\right), \quad (16)$$

where $\Gamma$ is the gamma function. The results of the fits and the values for $\beta$ can be found in Table S1. Finally, the electron transverse relaxation of the TEMPOL solution at 77 K and 1.2 T was investigated using the Carr-Purcell-Meiboom-Gill (CPMG) pulse train with a 2-step phase cycle as shown in Figure 6e. The echo delay d4 was 390 ns. Similarly, the echo detected signal was integrated and plotted to extract the characteristic time constant $T_{2S}$ reflecting the spin-spin relaxation time. The decaying stretched exponential function:

$$I_t^{\text{EPR}} = I_\infty^{\text{EPR}} \left(\exp\left(-\frac{t}{T_{2S}}\right)^\beta\right) \quad (17)$$

was used to fit the experimental data and the average relaxation time constant was calculated using equation 16. The relaxation results can be found in Table S1.

### 4.3 DNP simulations

For interpretation of the $^1$H and $^{13}$C DNP spectra for trityl OX063 and TEMPOL radicals, we performed DNP spectra simulations based on calculating the extent of saturation that is caused by first- and second-order saturation of electron bins of the polarizing radical in an amorphous frozen solution. The model is adapted from previously published strategies by Hovav et al.[23] and Wenckebach[38] and in our case is based on solving equation 4 of electron saturation under microwave irradiation in time also considering spectral diffusion. The partial differential equation can be solved over time using discretization of the problem with computer-based numerical approaches (MATLAB, ODE15s). However, when the dynamics of the electron saturation are being much faster than the typical continuous wave saturation time of the microwave, a steady-state electron polarization gradient can be assumed. At 1 T, the typical DNP buildup times are on the second timescale, while electron saturation happens after a few milliseconds. In that case, a time-independent analytical solution to the master equation 4 can also be used as an approximation to predict the hole burning shape. First, we must assume that microwave irradiation is long enough such that a steady state is reached after $t_{inf}$ and that electron polarization cannot exceed Boltzmann polarization:

$$\frac{\partial P_S(\omega_S, t_{\text{inf}})}{\partial t} = 0, P_S(\omega_S, t_{\text{inf}}) \leq P_L. \quad (18)$$

Moreover, when the contribution of direct saturation due to on-resonant microwave irradiation is negligible compared to the effect of spectral diffusion on the total electron saturation degree, the general form of equation 4 can be simplified to the following differential equation[19]:

$$\mathfrak{D}\frac{\partial^2 P_S(\omega_S)}{\partial \omega_S^2} = \delta(\omega_s - \omega_m)P_S(\omega_S) + \frac{1}{T_{1S}}(P_S(\omega_S) - P_L). \quad (19)$$

In other words, when homogeneous lineshape of the electron is much smaller than the inhomogeneous linewidth, $h(\omega_S - \omega_m)$ in equation 3 can be replaced with a delta function that is only zero at $\omega_s = \omega_m$ and equal to 1 when $\omega_s \neq \omega_m$. In case of 50 mM TEMPOL, the homogeneous linewidth of the electron is approximately $1/T_{2S}$ or 1.23 MHz when taking $T_{2S}$ = 0.81 μs (Table 2) if considered a Lorentzian shape. In contrast, the inhomogeneous linewidth stretches over more than 200 MHz at 1 T as can be seen in Figure 4b. A solution to equation 19 is represented in equation 8 and can be used as an analytical approach to calculate the effect of the microwave on the saturation of electrons in broad EPR spectra:

$$P_S(\omega_s) = P_L\left[1 - s \cdot \exp\left(-\frac{|\omega_S - \omega_m|}{\sqrt{\mathfrak{D}T_{1S}}}\right)\right], \quad (8)$$

where $s$ is the saturation factor. The saturation factor can only be estimated properly if solved numerically, but nevertheless it does not change the shape of the resulting DNP spectra when applying the SE and CE matching conditions in equation 9 and 11. All the codes used for simulation of the DNP spectra are made available in the repository.[42]

## 5. Author Contributions

SJ and EV conceived the project. EV and CB performed the experimental work. SVK, SAJ and GM provided the Q-band CW and pulse EPR measurements on TEMPOL and were important for interpretation of the results. JK, DB and RM provided the prototype benchtop DNP polarizer and further technological and scientific support. EV supervised the work that was performed, analysed the data, wrote, and ran the DNP simulation code, and wrote the manuscript. QS reviewed the mathematical derivations and was key for constructing the electron saturation code. All authors contributed to refining the manuscript.

## 6. Conflicts of interest

There are no conflicts to declare. The authors declare no competing financial interest. JK, DB and RM are employees of Bruker who co-developed the benchtop 1 T DNP polarizer system.

## 7. Acknowledgements

We acknowledge Bruker Biospin for providing the prototype benchtop DNP polarizer. We thank Tom Wenckebach and Asif Equbal for the fruitful discussions regarding the DNP mechanisms and simulations in benchtop and high temperature conditions. We additionally acknowledge C. Jose and C. Pages for use of the ISA Prototype Service, and S. Martinez of the UCBL mechanical workshop for machining parts of the experimental apparatus. This research was supported by, ENS-Lyon, the French CNRS, Lyon 1 University, the Deutsche Forschungsgemeinschaft (Grant No: SFB 1527, Project No. 454252029), the European Research Council under the European Union's Horizon 2020 research and innovation program (ERC Grant Agreements No. 101044726/HypFlow) and the French National Research Agency (project 'HyMag' ANR-18-CE09-0013).

# Supporting information

# Dynamic Nuclear Polarization Mechanisms using TEMPOL and trityl OX063 radicals at 1 T and 77 K

Ewoud Vaneeckhaute*[1], Charlotte Bocquelet[1], Nathan Rougier[1], Shebha Anandhi Jegadeesan[2], Sanjay Vinod-Kumar[2], Guinevere Mathies[2], Roberto Melzi[3], James Kempf[4], Quentin Stern[1], Sami Jannin[1]

*1. Université Claude Bernard Lyon 1, CNRS, ENS Lyon, UCBL, CRMN UMR 5082, 69100 Villeurbanne, France*

*2. Department of Chemistry, University of Konstanz, Universitätsstr. 10, 78464, Konstanz, Germany*

*3. Bruker Italia S.r.l., Viale V. Lancetti 43, 20158 Milano, Italy*

*4. Bruker Biospin, Billerica, Massachusetts 01821, United States*

*All the data and codes are available at* http://doi.org/10.5281/zenodo.14338967

*\* Corresponding author*

*Email: ewoud.vaneeckhaute@univ-lyon1.fr*

# 1. Supporting data

## 1.1. CW EPR of 50 mM TEMPOL at Q-band 1.2 T, 77 K

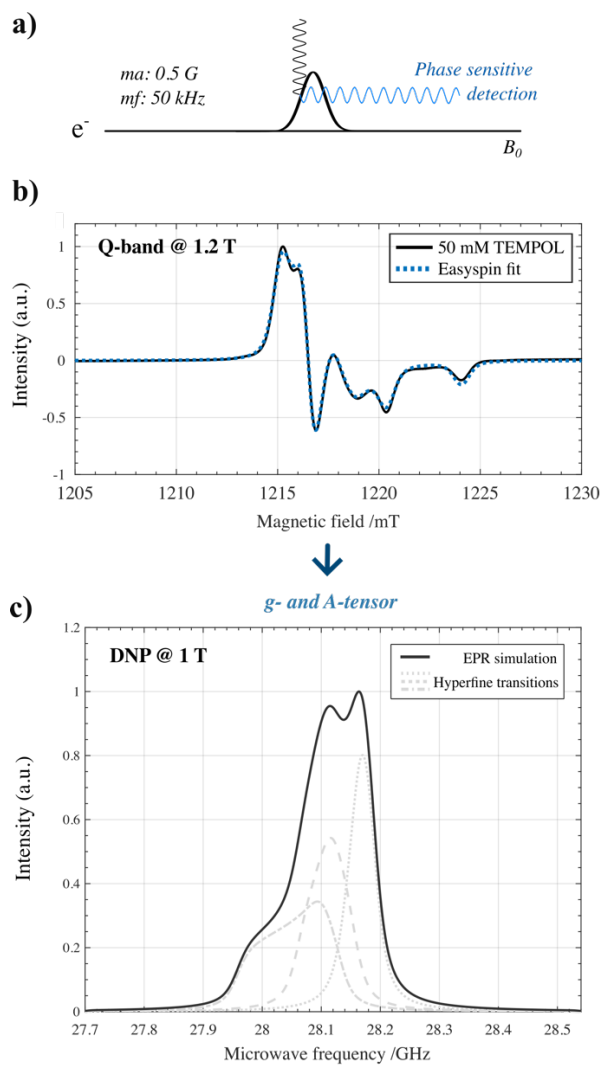

**Figure S 1.** a) Phase-sensitive continuous wave (CW) EPR experiment measured via sweeping the magnetic field with a field modulation amplitude (ma) of 0.5 G and a modulation frequency (mf) of 50 kHz. b) Q-band CW-EPR spectrum of 50 mM TEMPOL in a partially protonated 2:2:6 $H_2O:D_2O:DMSO\text{-}d_6$ (v:v:v) DNP solution measured at 1.2 T and 77 K. The Easyspin MATLAB package was used to extract the anisotropic g-tensor and hyperfine A-tensor values at 1.2 T. c) A simulated frequency-swept EPR spectrum of 50 mM TEMPOL to rationalize the EPR profile used for performing DNP at benchtop polarizer. An overlay of each contribution of the $^{14}N$ hyperfine transitions to the overall simulated DNP spectrum is visualized as well in dotted, striped, and dotted-striped lines. Fitted g-tensor and hyperfine A-tensor values for 50 mM TEMPOL extracted from the experimental CW-EPR spectrum are reported in Table 1 of the main manuscript.[1]

## 1.2. Pulse EPR versus CW EPR of 50 mM TEMPOL at Q-band 1.2 T, 77 K

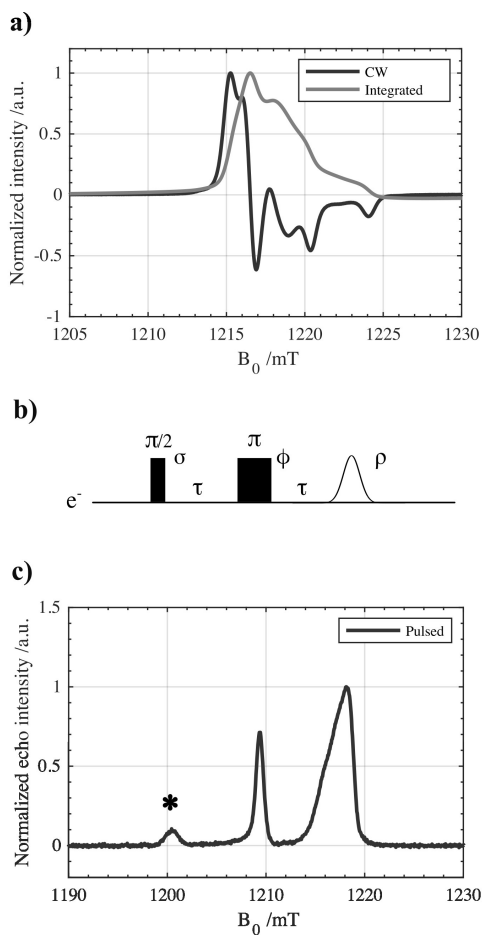

**Figure S 2.** Q-band CW field-sweep EPR spectrum (a) and echo-detected field-sweep pulse EPR spectrum (c) of 50 mM TEMPOL radical dissolved in a partially protonated 2:2:6 H$_2$O:D$_2$O:DMSO-d$_6$ (*v*:*v*:*v*) DNP solution measured at 77 K and 1.2 T. The pulse sequence used for the echo-detected field-sweep EPR spectrum is shown in (b) with π/2 = 14 ns and fixed echo time τ = 800 ns using a two-step phase cycle with σ = +x, +x and ϕ = +x, -x and with echo detection phase ρ = +, +. See the materials and methods section of the main manuscript for more detailed information on the EPR measurement conditions.

The presence of strong mutual electronic interactions in TEMPOL essential for cross effect DNP can be witnessed when comparing the integrated CW-EPR profile (grey) in Figure S2a with the echo-detected pulse EPR profile on a 50 mM TEMPOL frozen DNP solution in Figure S2c. A dramatic difference in the resulting EPR line shapes is apparent. The reason for the discrepancy in line shape can be explained by the fast spin-spin ($T_{2S}$) relaxation time constants rivalling the echo time τ of 800 ns observed at 77 K and high radical concentration of 50 mM (*vide infra*).

## 1.3. Electron spin-lattice and spin-spin relaxation of 50 mM TEMPOL at Q-band 1.2 T, 77 K

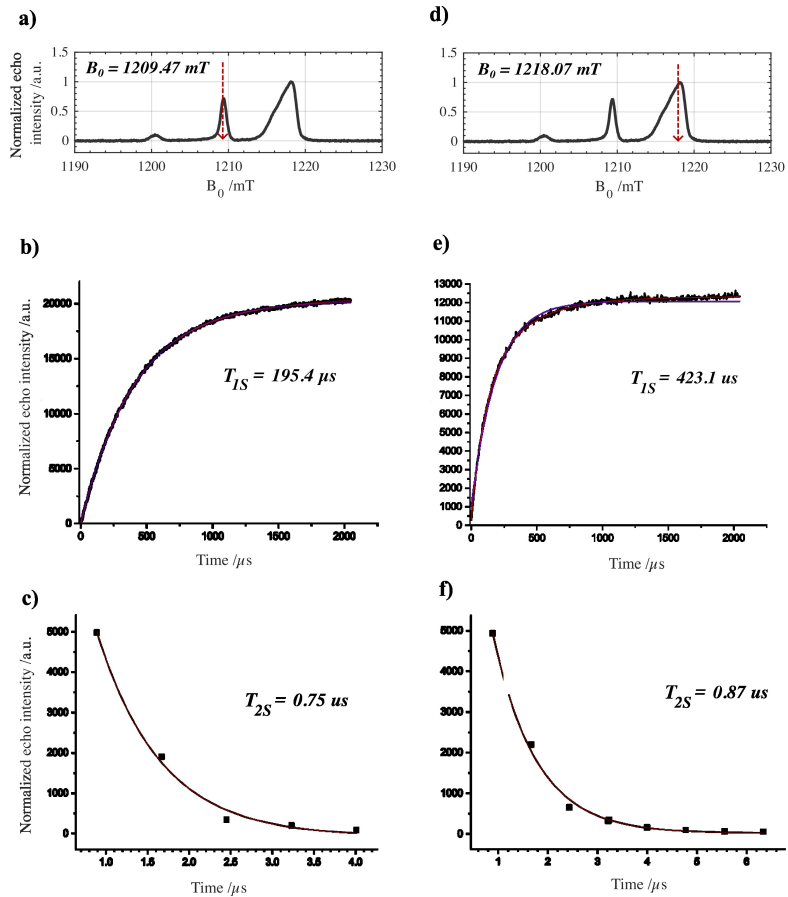

**Figure S 3.** Electron relaxation properties of 50 mM TEMPOL in a partially protonated 2:2:6 $H_2O:D_2O:DMSO-d_6$ (*v:v:v*) DNP solution measured at 77 K and Q-band at 1.2 T at different magnetic field positions of 1209.47 mT (a) and 1218.07 mT (d). Underneath, the electron spin-lattice relaxation times along the EPR line were measured for the frozen DNP solution using the saturation recovery pulse train shown in Figure 6d of the main manuscript. Spin-lattice relaxation constants were extracted using a stretched exponential at two different microwave frequencies visualized in (b,e). The stretch parameter fits are shown in Table S1. Also the empirical electron spin-spin relaxation times along the EPR line were measured using the CPMG-based pulse train shown Figure 6e. Analogous, spin-spin relaxation constants were extracted using the same procedure and shown in Table S1. See the materials and methods section of the main manuscript for more detailed information on the EPR measurement conditions.

$T_{1S}$ ranges from 195 µs to 423 µs depending on the position in the EPR profile. Several previous research groups have also observed different relaxation properties depending on the position in the EPR profile[2–6], which once again illustrates the presence of different inhomogeneous spin populations in case of broad-line TEMPOL. Different $T_{2S}$ spin-spin relaxation times are also apparent for different spin populations ranging from 0.75 µs to 0.87 µs. The fast decoherence of in-plane magnetization starts to rival with the implemented echo time of 0.8 µs in Figure S2b and therefore creates artifacts in the echo-detected pulse EPR as it is no longer possible to detect the electron spin populations. A similar phenomenon was also observed in case of high narrow-line trityl radical concentrations when measuring echo-detected EPR profiles.[5]

Table S1. Fitted values for the electron spin-lattice and spin-spin relaxation times for 50 mM TEMPOL dissolved in 2:2:6 $H_2O:D_2O:DMSO-d_6$ (v:v:v) solution employing a stretched exponential to the pulsed EPR measurements performed at Q-band (1.2 T, 34 GHz). A stretch coefficient $\beta < 1$ reflects a distribution of relaxation time constants. The average $T_{1S,av}$ can then be calculated using the gamma function using equation 16 in the materials and methods section.

| [TEMPOL] | Spin-lattice | | Spin-spin | |
|---|---|---|---|---|
| | $T_{1S,av}$ (µs) | $\beta$ | $T_{2S,av}$ (µs) | $\beta$ |
| 50 mM | 195.4 | 0.81 | 0.75 | 1.00 |
| | 423.1 | 0.98 | 0.87 | 1.00 |

## 1.4. Numerical versus analytical solution for simulating electron depolarization under microwave perturbation

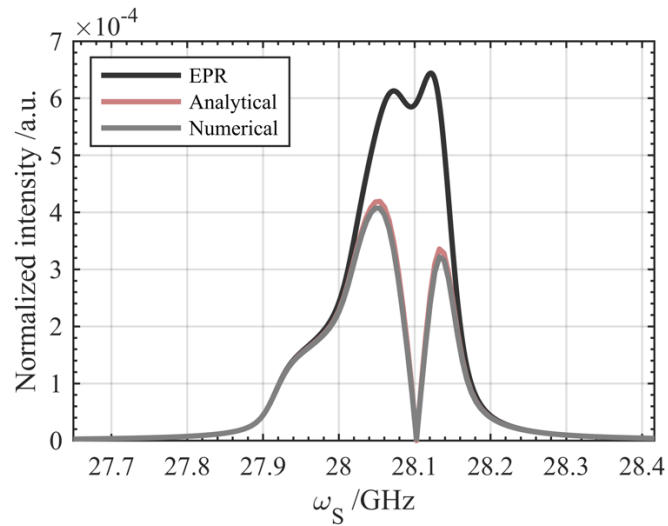

Figure S 4. Comparison of the analytical solution for approximating the hole burn shape in TEMPOL (equation 8 of the main manuscript) and the numerical solution to the master equation 4 for describing the dynamics of electron saturation under microwave irradiation. The values for the simulations are summarized in Table 2 of the main manuscript.

## 1.5. Spontaneous polarization transfer from protons to carbons in acetate at 1 T and 77 K using trityl OX063

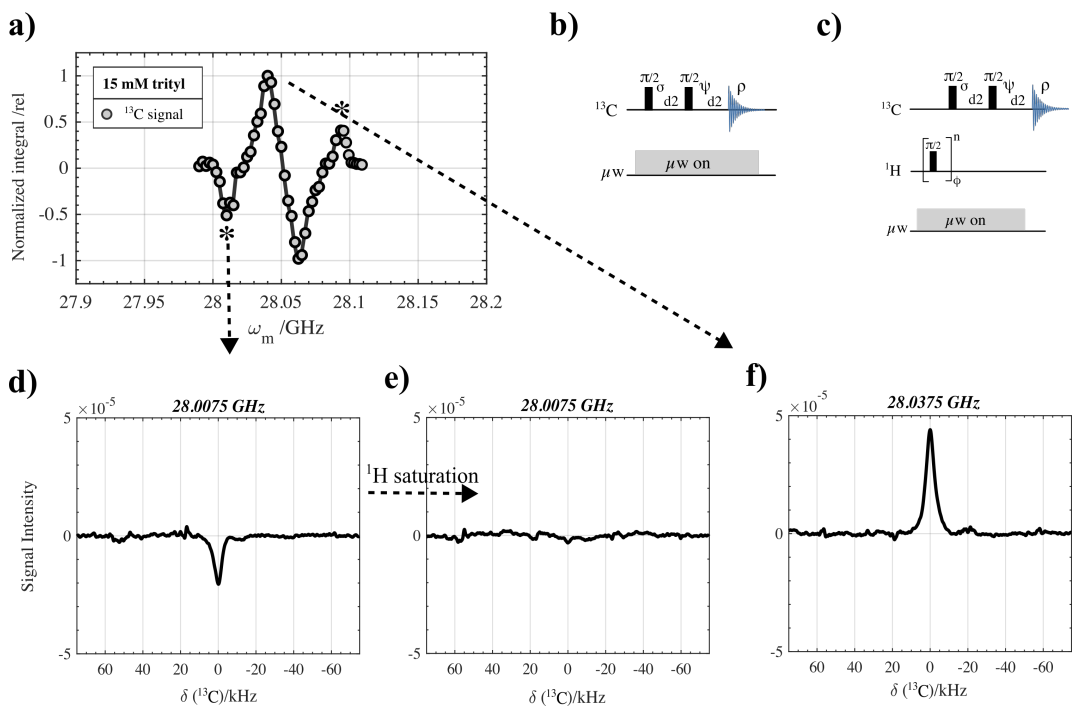

**Figure S 5.** Investigation of the additional $^{13}$C DNP optima (asterisk) with reversed polarization occurring when using 15 mM trityl OX063 with 3 M fully labelled [1-$^{13}$C] sodium acetate. a) Experimental $^{13}$C DNP spectra of acetate dissolved in 2:2:6 H$_2$O:D$_2$O:DMSO-d$_6$ (*v:v:v*) DNP solution. b-c) 1D $^{13}$C solid echo pulse sequence with and without saturation of protons during the DNP process. The echo delay d2 was set to 14 µs in between the center of the π/2 pulses and an 8-step phase cycling was used with σ = +x, -y, -x, y, +x, -y, -x, y and ψ = -y, -x, y, x, y, x, -y, -x and with detection phase ρ = -y, -x, y, x, -y, -x, y, x. d) Solid-state hyperpolarized 1D $^{13}$C spectrum (negative polarization) of acetate when irradiating at 28.0075 GHz without proton saturation during buildup. e) Solid-state hyperpolarized 1D $^{13}$C spectrum (no polarization) when irradiating at 28.0075 GHz with proton saturation during buildup. e) Solid-state hyperpolarized 1D $^{13}$C spectrum (positive polarization) when irradiating at 28.0375 GHz during buildup.